\documentclass[12pt]{article}
\usepackage{amsmath,amssymb,amsfonts,amsthm}

\title{Entanglement analysis of two-atom nonlinear JCM with nondegenerate two-photon transition, Kerr nonlinearity and two-mode Stark shift}

\author{H R Baghshahi$^{1,2,3}$, M K Tavassoly$^{1,2,*}$ and M J Faghihi$^{4}$ \\
 \footnotesize{$^1$ Atomic and Molecular Group, Faculty of Physics, Yazd University, Yazd, Iran} \\
 \footnotesize{$^2$ The Laboratory of Quantum Information Processing, Yazd University, Yazd, Iran} \\
 \footnotesize{$^3$ Department of Physics, Faculty of Science, Vali-e-Asr University of Rafsanjan, Rafsanjan, Iran} \\
 \footnotesize{$^4$ Physics and Photonics Department, Graduate University of Advanced Technology, Mahan, Kerman, Iran} \\
 \footnotesize{$^*$ E-mail: mktavassoly@yazd.ac.ir}}

\begin{document}
\maketitle

 \newcommand{\norm}[1]{\left\Vert#1\right\Vert}
 \newcommand{\abs}[1]{\left\vert#1\right\vert}
 \newcommand{\set}[1]{\left\{#1\right\}}
 \newcommand{\R}{\mathbb R}
 \newcommand{\I}{\mathbb{I}}
 \newcommand{\C}{\mathbb C}
 \newcommand{\eps}{\varepsilon}
 \newcommand{\To}{\longrightarrow}
 \newcommand{\BX}{\mathbf{B}(X)}
 \newcommand{\HH}{\mathfrak{H}}
 \newcommand{\A}{\mathcal{A}}
 \newcommand{\D}{\mathcal{D}}
 \newcommand{\N}{\mathcal{N}}
 \newcommand{\x}{\mathcal{x}}
 \newcommand{\p}{\mathcal{p}}
 \newcommand{\la}{\lambda}
 \newcommand{\af}{a^{ }_F}
 \newcommand{\afd}{a^\dag_F}
 \newcommand{\afy}{a^{ }_{F^{-1}}}
 \newcommand{\afdy}{a^\dag_{F^{-1}}}
 \newcommand{\fn}{\phi^{ }_n}
 \newcommand{\HD}{\hat{\mathcal{H}}}
 \newcommand{\HDD}{\mathcal{H}}

 \begin{abstract}
Entangled state, as an essential tool in quantum information processing, may be generated through the interaction between light and matter in cavity quantum electrodynamics. In this paper, we study the interaction between two two-level atoms and a two-mode field in an optical cavity enclosed by a medium with Kerr nonlinearity in the presence of detuning parameter and Stark effect. It is assumed that atom-field coupling and third-order susceptibility of the Kerr medium depend on the intensity of light. In order to investigate the dynamics of the introduced system, we obtain the exact analytical form of the state vector of the considered atom-field system  under initial conditions which may be prepared for the atoms (in a coherent superposition of their ground and upper states) and the fields (in standard coherent state). Then, in order to evaluate the degree of entanglement between subsystems, we investigate the dynamics of entanglement through the well-known criteria such as von Neumann reduced entropy, entanglement of formation and negativity.
Finally, we analyze the influences of Stark shift, deformed Kerr medium, intensity-dependent coupling and also detuning parameter on the above-mentioned measures, in detail. Numerical results show that the amount of entanglement between different subsystems can be controlled by choosing the evolved parameters, appropriately.
 \end{abstract}


 \section{Introductory remarks}\label{sec-intro}
The notion of entanglement, as the nonlocality aspect of quantum correlations, is a form of quantum superposition and an outstanding trait of quantum mechanics which has no classical counterpart; the concept that is known as the heart of the Einstein-Podolsky-Rosen (EPR) paradox \cite{Einstein1935} and Bell's theorem \cite{Bell1964}. The entanglement is an essential ingredient and a cornerstone of quantum information science such as quantum computation and communication \cite{Bennett.DiVincenzo2000}, quantum dense coding \cite{Barzanjeh.etal2013}, quantum teleportation \cite{Ganguly.etal2011}, entanglement swapping \cite{Branciard.etal2012}, sensitive measurements \cite{Kitagawa.etal2011}, quantum telecloning \cite{Koike.etal2006}, quantum key distribution \cite{Leverrier.etal2013} and quantum cryptography \cite{Jennewein.etal2000}.
Construction and generation of entangled states have attracted a great deal of attention in recent years. In particular, the appearance of entanglement in the light-matter interaction in cavity quantum electrodynamics (QED), which is regarded as a simple way to generate the entangled states, is of special interest. It has been reported that, the atom-field entangled states have been experimentally generated via a single atom interacting with a mesoscopic field in a high-Q microwave cavity \cite{Auffeves2003}. \\
%
The full quantum mechanical approach to the study of two-level atom interacting with a single-mode quantized field in the electric-dipole and rotating wave approximations
was first introduced via the well-known Jaynes-Cummings model (JCM) \cite{Jaynes.Cummings1963,Cummings1965}.
Many generalizations of the JCM have been proposed in various ways, for instance, considering different initial conditions \cite{Kuklifmmodecutenlseniski.Madajczyk1988}, entering the effects of dissipation and damping in the model \cite{rodriguez2005combining}, adopting multi-level atom \cite{Faghihi2014,Faghihi2014a,Baghshahi.etal2014CTP} as well as multi-photon transition \cite{Baghshahi.etal2014} and multi-atom \cite{Bougouffa1,Bougouffa2}. The dependence of atom-field coupling on the intensity of light is also considered as another generalization of the JCM which was suggested by Buck and Sukumar  \cite{Buck.Sukumar1981,Buck.Sukumar1984} and then was used by others \cite{Tavassoly.Yadollahi2012,Faghihi.Tavassoly2013,Baghshahi.Tavassoly2014}. In detail, quantum properties of a $\Lambda$-type three-level atom interacting with a single-mode field in a Kerr medium with intensity-dependent coupling and in the presence of the detuning parameters have been studied by us \cite{Faghihi.Tavassoly2012}.
Also, the authors discussed the nonlinear interaction between a three-level atom (in a $\Lambda$ configuration) and a two-mode cavity field in the presence of a cross-Kerr medium and its deformed counterpart \cite{Honarasa.Tavassoly2012}, intensity-dependent atom-field coupling and the detuning parameters \cite{Faghihi.etal2013,Faghihi.Tavassoly2013a}.\\
From another perspective of this field of research, multi-photon transitions in the JCM may be taken into account. Multi-photon process in atomic systems has attracted a great deal of attention, since this phenomenon results in the high degree of correlation between emitted photons which may lead to the nonclassical behavior of emitted light \cite{Rudolph.etal1998}. The concept of multi-photon transition can be more clarified after considering the Stark shift. The importance of this effect will be apparent, when two atomic levels are coupled with comparable strength to the intermediate level \cite{Alsing.Zubairy1987}. In this case, by applying the method of adiabatic elimination of the intermediate level(s) of a multi-level atom, the Stark shift phenomenon is revealed \cite{Swain1994}. Indeed, by using this method, one arrives at a simple  effective Hamiltonian in which a two-level atomic system interacts with a single-mode quantized field with multi-photon (at least two-photon) transition in the presence of Stark shift \cite{Brune.etal1987a}. It is instructive to state that, the Stark shift in two-photon transitions can be regarded as an intensity-dependent detuning \cite{Agarwal.Puri1986}. \\
In particular and in direct relation to the present work, the state vector of the system containing the interaction between two identical two-level atoms and a two-mode quantized radiation field via nondegenerate two-photon transition has been explicitly found in \cite{Bashkirov2006}. In this attempt, a few nonclassicality features such as photon statistics and squeezing have been numerically discussed. An exact solution for two two-level atoms interacting with a two-mode radiation field containing nondegenerate two-photon and Raman transitions has been proposed \cite{Bashkirov2008}, in which the amount of atom-field entanglement via the reduced atomic entropy has been examined. The interaction between two two-level atoms and a single-mode field with degenerate two-photon transition in the presence of the Stark shift has been studied in \cite{Obada2013c} in which the authors showed that the degree of entanglement (DEM) may be improved after increasing the value of the Stark shift parameter. Recently, dynamical behavior of two two-level atoms interacting with a single-mode binomial field has been studied by one of us \cite{Hekmatara2014}, in which the role of the parameter related to the dimension of the binomial state on various dynamical properties such as the atomic population inversion, sub-Poissonian statistics and also entropy squeezing for Pauli operators have been evaluated.\\
In this paper, we intend to outline the nonlinear interaction between two two-level atoms and a two-mode quantized cavity field within an optical cavity surrounded by a centrosymmetric medium with Kerr nonlinearity (containing its $f$-deformed counterpart) in the presence of detuning parameter and Stark effect. Our main goal is to investigate the effects of these parameters, namely intensity-dependent coupling, Kerr (and also deformed Kerr) medium (containing self- and cross-action), Stark shift and detuning parameter on the temporal behaviour of entanglement of formation (EOF) measure.\\
To make our motivations more clear, a few words on the notability of the introduced model should be given. In this respect, it has been shown that, from the standpoint of the theory of quantum information, the atomic systems can be regarded as an inherent computational hardware that is necessary for the future implementation of quantum information protocols \cite{Leibfried.etal2003,Julsgaard.etal2004}. In addition, the photons of the field can be considered as a fundamental building blocks of quantum communication \cite{Bouwmeester.etal1997} and cryptography \cite{Gisin.etal2002}. Accordingly, it may be expressed that, generating and manipulating the nonclassical correlations arisen from the interaction between atomic systems and quantized radiation fields, place this kind of interaction at the frontier subjects of the field of quantum information. In particular case related to the atomic systems where two atoms participate in the interaction, it is valuable to state that, since a two-level atom can represent a qubit, the atomic system consisting of two (identical) two-level atoms (two qubits) can be applied as a quantum gates in protocols in quantum information processing. In this relation, it is remarkable to state that, two-atom entangled states have experimentally been realized via ultra cold trapped ions \cite{DeVoe1996} and cavity QED schemes \cite{Zheng.Guo2000}.\\
 The remainder of paper is organized as follows: In the next section, the analytical form of the state vector of the whole system is obtained. Section \ref{Entanglement of formation} deals with two different regimes of entanglement (atom-field and atom-atom entanglement) by evaluating the EOF measure.
 Finally, section \ref{Summary} contains a summary and concluding remarks.
%
 \section{Introducing the model Hamiltonian  and its solution}\label{Model}
%
This section is devoted to finding the explicit form of the state vector of the system, since based on the fundamentals of quantum mechanics, possible information during studying any physical (quantum) system is hidden in its wave function. For this purpose, it is necessary to take all interactions between subsystems into account by suitably using the fully quantum mechanical approach. Then, with the help of the Schr\"{o}dinger equation or other appropriate (equivalent) methods, the state vector of the whole system may be found. So, let us consider a model in which a two-mode quantized radiation field oscillating with frequencies $\Omega_{1}$ and $\Omega_{2}$ interacts with two two-level atoms (atom $A$ and atom $B$) with ground states $|g_{A}\rangle$, $|g_{B}\rangle$ and exited states $|e_{A}\rangle$, $|e_{B}\rangle$, in an optical cavity surrounded by a centrosymmetric medium with Kerr nonlinearity in the presence of the Stark shift and detuning parameter. Also, considering the centrosymmetric nonlinear medium, self- and cross-action of the Kerr nonlinearity are properly taken into account. In addition, being in the intensity-dependent regime, it is assumed that atom-field coupling and nonlinear susceptibility of Kerr medium are $f$-deformed (see figure \ref{Diagram}). Anyway, keeping in mind the fact that, the transition between the atomic levels is accompanied by the absorption/emmision of two photons, the Hamiltonian comprising all above-mentioned interactions which describes the dynamics of the introduced physical system in the RWA can be written as ($\hbar=c=1$):
 \begin{eqnarray} \label{Phy 1}
 \hat{H}&=& \sum_{i=A,B}\omega_{i} \hat{\sigma}_{z}^{(i)}+\sum_{j=1}^{2} \Omega_{j} \hat{a}_{j} ^{\dag} \hat{a}_{j}+ \sum_{i=A,B} \lambda_{i} \left( \hat{A}_{1} \hat{A}_{2} \hat{\sigma}_{+}^{(i)}  +  \hat{A}_{1} ^{\dag} \hat{A}_{2} ^{\dag} \hat{\sigma}_{-}^{(i)} \right )   \nonumber \\
&+& \sum_{j=1}^{2} \chi_{j} \hat{R}_{j}^{\dagger^{2}} \hat{R}_{j}^{2} +\chi \hat{R}_{1}^{\dagger} \hat{R}_{1} \hat{R}_{2}^{\dagger} \hat{R}_{2}   \nonumber \\
&+&  \sum_{i=A,B} \left( \beta_{1}^{(i)} \hat{a}_{1} ^{\dag} \hat{a}_{1} \hat{\sigma}_{-}^{(i)} \hat{\sigma}_{+}^{(i)} +\beta_{2}^{(i)} \hat{a}_{2} ^{\dag} \hat{a}_{2}\hat{\sigma}_{+}^{(i)} \hat{\sigma}_{-}^{(i)} \right ).
\end{eqnarray}
In above relation, $\hat{\sigma}_{z}^{(i)}$ and $\hat{\sigma}_{\pm}^{(i)}$  are the atomic pseudospin operators for the $i$th atom, $\hat{a}_{j}$ ($\hat{a}_{j} ^{\dag}$) is the bosonic annihilation (creation) operator of the field mode $j$, $\omega_{i}$ shows the frequency of atomic transition, $\lambda_{i}$ is related to the atom-field coupling constant, and $\beta_{1}^{(i)}$ and $\beta_{2}^{(i)}$ are the effective Stark shift coefficients. Also, $\chi_{j}$ and $\chi_{12}$ denote the cubic susceptibility of the medium; $\chi_{j}$ represents the Kerr self-action for mode $j$, while $\chi_{12}$ is related to the Kerr cross-action process. In addition, $\hat{A}_{j}=\hat{a}_{j} f(\hat{n}_{j})$ and $\hat{R}_{j}=\hat{a}_{j} g(\hat{n}_{j})$ are the nonlinear ($f$-deformed) annihilation operators with $\hat{A}_{j}^{\dag}$ and $\hat{R}_{j}^{\dag}$ as their respective Hermitian conjugates, where $\hat{n}_{j}= \hat{a}_{j} ^{\dag} \hat{a}_{j}$ and $f(\hat{n}_{j})$ and $g(\hat{n}_{j})$ correspond to the Hermitian operator-valued functions responsible for the intensity-dependent atom-field coupling and $f$-deformed Kerr nonlinearity. These operators satisfy the communication relations $[\hat{A}_{j},\hat{n}_{j}]=\hat{A}_{j} $,  $[\hat{A}_{j}^{\dag},\hat{n}_{j}]=-\hat{A}_{j}^{\dag}$,  $[\hat{R}_{j},\hat{n}_{j}]=\hat{R}_{j}$ and $[\hat{R}_{j}^{\dag},\hat{n}_{j}]=-\hat{R}_{j}^{\dag}$.
By comparing the Hamiltonian in (\ref{Phy 1}) with the standard JCM it is seen that, we have in fact made the transformations $\lambda_{i} \rightarrow \lambda_{i} f_{1}(\hat{n}_{1}) f_{2}(\hat{n}_{2})$, $\chi_{j} \rightarrow \chi_{j} g_{j}^{2}(\hat{n}_{j}) g_{j}^{2}(\hat{n}_{j}-1)$ and $\chi \rightarrow \chi g_{1}^{2}(\hat{n}_{1}) g_{2}^{2}(\hat{n}_{2})$ (the physical motivation of this typical change has been established in \cite{Honarasa.Tavassoly2012}).  In order to arrive at a universal formalism for the considered atom-field system, the intensity-dependent functions are applied in the general form $f_{j}(\hat{n}_{j})$ and $g_{j}(\hat{n}_{j})$. Accordingly, it is worth mentioning that selecting various nonlinearity functions leads to the different Hamiltonians and as a result, different state vectors may be obtained.      \\
Looking deeply at the relation (\ref{Phy 1}) implies the fact that, the introduced Hamiltonian signifies the intensity-dependent two-atom two-mode two-photon JCM in the presence of a $f$-deformed Kerr nonlinearity and the Stark shift. It is valuable to express that, the third part in Hamiltonian (\ref{Phy 1}) indicates the effect of two-mode Stark shift \cite{Gou1990}, which may be interpreted as the intensity-dependent energy shifts of the atomic levels \cite{Agarwal.Puri1986}.
It is noteworthy to state that, the process of two-mode two-photon transition can be physically demonstrated via non-degenerate two-photon process, in which the atom makes a transition from its ground (excited) state to an excited (ground) state by the simultaneous absorption (emission) of two laser photons \cite{Boyd2008}. Indeed, in the case of two-photon absorption, for instance, the atom first absorbs a photon of frequency $\Omega_{1}$ and jumps from a {\it real} level to the higher {\it virtual} one, and then by absorbing a photon of frequency $\Omega_{2}$, jumps to the {\it nearest} real level. \\
Anyway, for the next purpose, it is suitable to rewrite the Hamiltonian (\ref{Phy 1}) in the interaction picture from which one arrives at
\begin{eqnarray} \label{Phy 4}
 \hat{H_{I}}&=& \sum_{i=A,B} \lambda_{i} \left( \hat{A}_{1} \hat{A}_{2} \hat{\sigma}_{+}^{(i)} e^{i\Delta_{i} t} +  \hat{A}_{1} ^{\dag} \hat{A}_{2} ^{\dag} \hat{\sigma}_{-}^{(i)} e^{-i\Delta_{i} t} \right )\nonumber\\
&+& \sum_{j=1}^{2} \chi_{j} \hat{R}_{j}^{\dagger^{2}} \hat{R}_{j}^{2} +\chi \hat{R}_{1}^{\dagger} \hat{R}_{1} \hat{R}_{2}^{\dagger} \hat{R}_{2}
\nonumber\\
&+&  \sum_{i=A,B} \left( \beta_{1}^{(i)} \hat{a}_{1} ^{\dag} \hat{a}_{1}\hat{\sigma}_{-}^{(i)} \hat{\sigma}_{+}^{(i)} +\beta_{2}^{(i)} \hat{a}_{2} ^{\dag} \hat{a}_{2}\hat{\sigma}_{+}^{(i)} \hat{\sigma}_{-}^{(i)} \right ),
\end{eqnarray}
where $\Delta_{i}=\omega_{i}-\Omega_{1}-\Omega_{2}$ is the detuning parameter.
Now, for simplicity and in order to be more realistic, let us consider the atoms to be identical, that is, $\Delta_{A} = \Delta_{B} \equiv \Delta$, $\beta_{1}^{(A)} = \beta_{1}^{(B)} \equiv \beta_{1}$, $\beta_{2}^{(A)} = \beta_{2}^{(B)} \equiv \beta_{2}$, $\lambda_{A} = \lambda_{B} \equiv \lambda$.
At this stage, to solve the Hamiltonian of the atom-field system, there exist three different but equivalent methods, namely probability amplitudes, Heisenberg operators and the unitary time evolution operator approaches \cite{Scully.Zubairy1997}. Altogether, we use the method of probability amplitudes. Therefore, let us assume that the wave function $|\psi(t) \rangle$ associated with the entire system at any time $t$ is in the following form
 \begin{eqnarray} \label{Phy 6}
|\psi(t) \rangle &=& \sum_{n = 0}^{ +\infty}  \sum_{m = 0}^{ +\infty} q_{n} q_{m} \Bigg[A(n,m,t) |e_{A},e_{B},n,m \rangle \nonumber\\
 &+& B(n+1,m+1,t) |e_{A},g_{B},n+1,m+1 \rangle \nonumber \\
 &+& C(n+1,m+1,t) |g_{A},e_{B},n+1,m+1 \rangle \nonumber \\
 &+& D(n+2,m+2,t) |g_{A},g_{B},n+2,m+2 \rangle \Bigg],
  \end{eqnarray}
where $q_{n}$ and $q_{m}$ are the probability amplitudes of the initial state of the radiation field of the cavity. Clearly, $A$, $B$, $C$ and $D$ are the time-dependent atomic probability amplitudes which have to be evaluated. Considering the probability amplitude technique, and after some lengthy but straightforward manipulations, the atomic probability amplitudes $A$, $B$, $C$ and $D$ (leading to the explicit form of the wave function of the whole system) are given by
\begin{eqnarray}\label{Phy 12}
A(n,m,t)=\sum_{j=1}^{3} \eta_{j} e^{i \xi_{j} t},    \nonumber  \\
B(n+1,m+1,t)=C(n+1,m+1,t)=\frac{-e^{-i \Delta t}}{2 k_{1}} \sum_{j=1}^{3}  (V_{A}+\xi_{j}) \eta_{j} e^{i \xi_{j} t},    \nonumber  \\
D(n+2,m+2,t) = \frac{e^{-2i \Delta t}}{2 k_{1} k_{2}} \sum_{j=1}^{3} \Bigg[(\xi_{j}+V_{A})(\xi_{j}-\Delta) \nonumber  \\
 \hspace{3.65cm}+ (\xi_{j}+V_{A})V_{B}-2 k_{1}^{2}  \Bigg] \eta_{j} e^{i \xi_{j} t},
\end{eqnarray}
where
\begin{eqnarray} \label{Phy 11}
\xi_{j}&=&-\frac{1}{3}x_{1} +\frac{2}{3} \sqrt{x_{1}^{2}-3x_{2}} \cos \Bigg[ \theta+\frac{2}{3} (j-1) \pi\Bigg],            \nonumber\\
\theta&=&\frac{1}{3} \cos ^{-1} \Bigg[ \frac{9 x_{1} x_{2} -2 x_{1}^{3}-27 x_{3}}{2(x_{1}^{2} - 3 x_{2})^{3/2}} \Bigg],
\end{eqnarray}
with
\begin{eqnarray} \label{Phy 9}
x_{1}&=&V_{A}+V_{B}+V_{D}-3\Delta, \nonumber\\
x_{2}&=&(V_{B}-\Delta)(V_{D}-2\Delta)+V_{A}(V_{B}+V_{D}-3\Delta)-2(k_{1}^{2}+k_{2}^{2}), \nonumber\\
x_{3}&=&(2\Delta-V_{D}) \left( V_{A} (\Delta-V_{B})+2k_{1}^{2}\right)-2 V_{A} k_{2}^{2}.
\end{eqnarray}
In the above relations we have defined
\begin{eqnarray}
V_{A} = V_{1}+2\beta_{2}m, \nonumber \\
V_{B} = V_{2}+\beta_{1}(n+1)+\beta_{2}(m+1), \nonumber \\
V_{D} = V_{3}+2\beta_{1}(n+2).
\end{eqnarray}
with
\begin{eqnarray}
\hspace{-2cm}V(n,m) = \chi_{1} n (n -1) g_{1}^{2}(n) g_{1}^{2}(n - 1)+\chi_{2} m (m -1) g_{2}^{2}(m) g_{2}^{2}(m -1)
 + \chi n m g_{1}^{2} (n) g_{2}^{2} (m), \nonumber \\
\hspace{-2cm}V_{1} = V(n,m), \hspace{1cm} V_{2}=V(n+1,m+1), \hspace{1cm}  V_{3}=V(n+2,m+2), \nonumber \\
\hspace{-2cm}k_{j} = \lambda f_{1}(n+j) f_{2}(m+j) \sqrt{(n+j)(m+j)},  \hspace{.5cm} j=1,2.
\end{eqnarray}
Notice that, the coefficients $\eta_{j}$ are still unknown. They can be determined via determining the initial conditions of the atoms. Suppose that the atoms are initially prepared in the coherent superposition of the exited and ground states $|e_{A},e_{B}\rangle$ and $|g_{A},g_{B}\rangle$, respectively, i.e.,
\begin{equation} \label{Phy 5}
|\psi(t=0) \rangle_{\mathrm{atoms}} =\cos(\varphi/2) |e_{A},e_{B} \rangle + \sin(\varphi/2) |g_{A},g_{B} \rangle,
\end{equation}
where $ 0 \leq \varphi \leq \pi $, i.e. $A(0) = \cos(\varphi/2)$, $B(0) = 0 =C(0)$ and $D(0)=\sin(\varphi/2)$. This superposition is specially known as the Bell state (maximally entangled state) if one sets $\varphi=\frac{\pi}{2}$.  However, by considering the general value of $\varphi$, the following relation may be obtained:
\begin{equation} \label{20}
\hspace{-2cm} \eta_{j}=\frac{2 \sin(\varphi/2)k_{1}k_{2}+\cos(\varphi/2) \bigg(2k_{1}^{2}+(\xi_{k} + V_{A})(\xi_{l} + V_{A}) \bigg)}{\xi_{jk}\xi_{jl}},\hspace{1cm} j\neq k \neq l = 1,2,3,
\end{equation}
with $\xi_{jk} = \xi_{j} - \xi_{k}$. Consequently, the probability amplitudes $A$, $B$, $C$ and $D$ are explicitly derived and as a result, the exact form of the wave function of the whole system is analytically obtained. \\
It is now worthwhile to declare that, studying the nonclassicality features of the state vector can be achieved after specifying the amplitudes of the initial states of the field which may be considered as number, phase, coherent or squeezed state. However, since the coherent state (the laser field far above the threshold condition \cite{Scully.Zubairy1997}) is more accessible than other typical field states, we shall consider the fields to be initially in the coherent states
 \begin{eqnarray}\label{amplitude}
\hspace{-2cm} |\alpha_{1}, \alpha_{2} \rangle = \sum_{n=0}^{+\infty}  \sum_{m=0}^{+\infty} q_{n} q_{m} |n,m \rangle, \;\;\;\;q_{n} = \exp \left( -\frac{ |\alpha_{1}|^{2} }{2}\right) \frac{\alpha_{1}^{n}}{\sqrt{n!}}, \;q_{m} = \exp \left( -\frac{ |\alpha_{2}|^{2} }{2}\right) \frac{\alpha_{2}^{m}}{\sqrt{m!}},
 \end{eqnarray}
 in which $|\alpha_{1}|^{2}$ and $|\alpha_{2}|^{2}$ represent the mean photon number (intensity of light) of mode $1$ and $2$, respectively.
 It is worthwhile mentioning that, the obtained formalism can be applied for any physical system with arbitrary nonlinearity function. In this paper, we use the nonlinearity function $ f(n)  = \sqrt{n} $ (associated with the atom-field coupling) where its associated coherent state is arisen naturally from the Hamiltonian illustrating the interaction with intensity-dependent coupling between a two-level atom and a radiation field \cite{Agarwal.Singh1982,Eftekhari.Tavassoly2010,Safaeian.Tavassoly2011}.
 Experimental verification of this function has been recently reported in \cite{Fink.etal2008}.
In addition, the physical interest of the nonlinearity function $ g(n)  = 1/\sqrt{n}$ (related to the intensity-dependent nonlinear susceptibility) which has been derived by Man'ko {\it et al} \cite{Manko.etal1997} from the coherent states, and the corresponding nonlinear coherent states has been called by Sudarshan as harmonious states \cite{Sudarshan1993}. This function is a popular nonlinearity function which has been usually used in the contents of deformation of bosonic operators in quantum optics literature \cite{Tavassoly2008,Tavassoly2010,Piroozi.Tavassoly2012}.
Anyway, we are now in a position to study the nonclassical features of the introduced quantum system. Nonclassicality of the radiation field states which may be generated through the nonlinear coherent states technique \cite{Mancini1997PLA}, has obtained a great attention in various fields of research \cite{ManciniEPL2001}.
Among all nonclassical properties, which are of special interest to the field of quantum optics and quantum information processing, we now pay attention to evaluate the entanglement dynamics of the obtained state, due to the significant role of quantum entanglement in the implementation of quantum information processing devices \cite{Alber.etal2001}.
It has been shown that, there exist some suitable measures being well justified and mathematically tractable, for instance, EOF and distillation \cite{Wootters1998}, negativity \cite{Vidal2002}, von Neumann and relative entropies \cite{Vedral1997} and concurrence \cite{Uhlmann2000}, from which quantum entanglement can be quantified \cite{Horodecki2009}. In the next section, in order to obtain the DEM between subsystems, the dynamics of EOF (to understand the atom-field entanglement as well as the atom-atom entanglement) is numerically evaluated. In each case, the effect of intensity-dependent coupling, deformed Kerr nonlinearity, detuning parameter and Stark shift is examined, in detail.
 %
 \section{Entanglement of formation}\label{Entanglement of formation}
The EOF is the most meaningful and physically motivated measure of quantum entanglement which clarifies the minimal cost that is required to prepare a special quantum state in terms of EPR pairs \cite{Bennett.etal1996}. The EOF has recently attracted a great deal of attention; for instance, it plays an important role in quantum phase transition for various interacting quantum multi-component systems \cite{Wu.etal2004}, the capacity of quantum channels \cite{Pomeransky2003} and may significantly affect macroscopic properties of solids \cite{Ghosh.etal2003}. \\
It has been shown that, for pure states, the EOF measure is defined as the entropy of either of two subsystems $A$ and $B$  as the following form \cite{Wootters1998,Bennett.etal1996}
\begin{equation} \label{EOFpure}
\mathcal{E}_{F}(|\psi\rangle)=-\mathrm{Tr} (\hat{\rho}_{A} \ln \hat{\rho}_{A})=-\mathrm{Tr} (\hat{\rho}_{B} \ln \hat{\rho}_{B}),
\end{equation}
 where $\hat{\rho}_{A}$ ($\hat{\rho}_{B}$) is the reduced density operator of subsystem $A (B)$. The definition of EOF is extended to a mixed state $\hat{\rho}$ by using the convex-roof method as \cite{Uhlmann2000,Bennett.etal1996}
\begin{equation} \label{EOF2}
\mathcal{E}_{F}(\hat{\rho})=\min \sum_{i} p_{i} \mathcal{E}_{F} (|\psi\rangle _{i})
\end{equation}
where the minimum is taken over all possible pure-state decompositions with $0 \leq p_{i} \leq1$ and $\sum_{i} p_{i}=1$. The ensemble that satisfies the minimum value in (\ref{EOF2}) is known as the decomposition of the mixed state optimal $\hat{\rho}$. Therefore, in order to find the EOF for a mixed state, the fundamental problem is to determine the optimal decomposition. Specifically, the EOF for bipartite mixed states is given by \cite{Wootters1998}
\begin{equation} \label{EOF3}
\mathcal{E}_{F}(\mathcal{C})=h\left( \frac{1+\sqrt{1-\mathcal{C}^{2}}} {2} \right),
\end{equation}
where $h(x)$ is the binary entropy function defined by $h(x) = - x \ln x - ( 1 - x ) \ln( 1 - x) $ (Shannon entropy) and $\mathcal{C}$ is called the concurrence.
Since the function $\mathcal{E}_{F}(\mathcal{C})$ is monotonically increased for $ 0 \leq \mathcal{C} \leq 1 $, the concurrence can be considered as a quantity that can measure the quantum entanglement. Although, unlike the EOF that is a resource-based or information theoretic measure, the concurrence is not \cite{Wootters2001}. Anyway, the concurrence for a bipartite system with density matrix $\rho$ reads as
\begin{equation} \label{EOF4}
\mathcal{C}(\rho)=\max \left (0,2 \max [\lambda_{j}]- \sum_{j=1}^{4} \lambda_{j} \right),
\end{equation}
in which $\lambda_{j}$ denotes the square roots of the eigenvalues of the operator $\hat{\tilde{\rho}} \hat{\rho}$ with $\hat{\tilde{\rho}}$ showing the ``spin-flipped'' density matrix defined by
 \begin{equation} \label{EOF5}
\hat{\tilde{\rho}}=(\hat{\sigma}_{y} \otimes \hat{\sigma}_{y}) \hat{\rho}^{\ast} (\hat{\sigma}_{y} \otimes \hat{\sigma}_{y}),
\end{equation}
where $\hat{\rho}^{\ast}$ and $\hat{\sigma}_{y}$ are the Hermitian conjugate of $\hat{\rho}$ and the `$y$' Pauli matrix, respectively. It is valuable to mention that, the concurrence varies from $\mathcal{C}=0$ (separable state) to $\mathcal{C} = 1$ (maximally entangled state) and the case $ 0 < \mathcal{C} < 1 $ indicates partially entangled state. In the next subsections we are going to examine the DEM between different subsystems, namely atoms-fields (pure state) and atom-atom (mixed state), by using the equations (\ref{EOFpure}) and (\ref{EOF3}), respectively.
\subsection{Atom-field entanglement}\label{Atom-field entanglement}
In order to study the DEM between subsystems (atoms and fields) of the obtained state, the EOF is a good measure. Because of in this case the system is pure state, we use the definition of EOF for pure states as
\begin{equation} \label{Von 2}
\mathcal{E}_{F}(t) = - \mathrm{Tr}_{A(F)}(\hat{\rho}_{A(F)}(t)\ln\hat{\rho}_{A(F)}(t)),
\end{equation}
with $\hat{\rho}_{A(F)}(t)=\mathrm{Tr}_{F(A)}(|\psi(t)\rangle \langle\psi(t)|)$, which denoted the reduced density operator of the atoms (fields).
Considering the procedure of \cite{Faghihi.etal2013}, the equation (\ref{Von 2}) may be expressed as follows:
\begin{equation} \label{Von 3}
\mathcal{E}_{F}(t) = - \sum_{j=1}^{4} \zeta_{j} \ln \zeta_{j},
\end{equation}
where $\zeta_{j}$ represents the eigenvalues of the reduced density operator of the atoms, which is given by the Cardano's method as \cite{Childs2009}
\begin{eqnarray}
\zeta_{j} &=& -\frac{1}{3} \varrho_{1}+\frac{2}{3}\sqrt{\varrho_{1}^{2} - 3 \varrho_{2}} \cos \left(\varpi+\frac{2}{3}(j-1)\pi \right), \;\;\;\;\;\;\;  j = 1,2,3, \nonumber \\
\zeta_{4} &=& 0,
 \end{eqnarray}
with
\begin{equation} \label{27}
\varpi=\frac{1}{3}\cos^{-1}\left[ \frac{9 \varrho_{1}\varrho_{2}-2 \varrho_{1}^{3} - 27 \varrho_{3}}{2( \varrho_{1}^{2} - 3\varrho_{2})^{3/2}}\right],
\end{equation}
and
\begin{eqnarray}
\varrho_{1} &=& -\rho_{11}-2\rho_{22}-\rho_{44}, \label{28-1} \\
\varrho_{2} &=& -2\rho_{12}\rho_{21}-\rho_{14}\rho_{41}-2\rho_{24}\rho_{42}+2\rho_{22}\rho_{44} + \rho_{11}(2\rho_{22}+\rho_{44}), \label{28-2} \\
\varrho_{3} &=& 2\rho_{14}(\rho_{22}\rho_{41}-\rho_{21}\rho_{42})+\rho_{12}(\rho_{21}\rho_{44}-\rho_{24}\rho_{41}) + \rho_{11}(\rho_{24}\rho_{42}
-\rho_{22}\rho_{44}), \label{28-3}
\end{eqnarray}
where the relation (\ref{28-1}) clearly shows that the parameter $\varrho_{1}$ is precisely equal to $-1$ and the matrix elements of the atomic density operator are as follows:
\begin{eqnarray}\label{281}
\rho_{ij}(t)&=&\sum_{n=0}^{+ \infty}\sum_{m=0}^{+ \infty} \langle n,m,i|\psi(t)\rangle \langle \psi(t)| n,m, j\rangle, \hspace{1cm}
i,j = 1,2,3,4.
\end{eqnarray}
where we have set $|1\rangle = |e_{A},e_{B}\rangle$, $|2\rangle = |e_{A},g_{B}\rangle$, $|3\rangle = |g_{A},e_{B}\rangle$ and $|4\rangle = |g_{A},g_{B}\rangle$.
Figure \ref{EOF} shows the evolution of the EOF against the scaled time $\tau$ for initial mean photon numbers fixed at $|\alpha_{1}|^{2} = 10 = |\alpha_{2}|^{2}$ and two atoms prepared initially in the ground state ($\varphi=\pi$). The left plots concern with the absence of the intensity-dependent coupling $(f (n) = 1)$ while the right ones correspond to the intensity-dependent coupling  with nonlinearity function $f (n)  = \sqrt{n}$. In figure \ref{EOF}(a), which is assumed the resonance case ($ \Delta= 0 $), there is neither Kerr effect ($\chi = 0$) nor Stark shift ($\beta_{1} = \beta_{2} = 0$). Figure 2(b) shows the effect of the $f$-deformed Kerr nonlinearity ($\chi_{1}=\chi_{2} = \chi / 2 =0.4 \lambda, g(n) = 1/\sqrt{n}$) in exact resonance condition. The role of detuning parameter ($\Delta =10 \lambda$) and Stark shift in the absence of other parameters is shown in figures 2(c) and 2(d), respectively.

In detail, from figure \ref{EOF}(a) whereas Kerr effect and detuning are disregarded, a random behaviour for the time evolution of the EOF is observed in both constant and intensity-dependent coupling regime; it is seen that intensity-dependent coupling has no considerable effect in the DEM. This situation is repeated in figure \ref{EOF}(b) with the difference that the deformed Kerr medium (which is distinguishable from Kerr medium through the nonlinearity function $g(n) = 1/\sqrt{n}$) can improve the maximum amount of the DEM in the presence of the intensity-dependent coupling. Figure \ref{EOF}(c) refers to the effect of detuning parameter. It is beheld that, in the presence of detuning parameter, the amount of the EOF and so the DEM is increased. According to figure \ref{EOF}(d), one can observe that, the amount of entanglement is reduced in the constant coupling regime. Although, in the presence of intensity-dependent coupling and in contrast to figure \ref{EOF}(a), the DEM is retained. \\
In summary, comparing the left plots of figure \ref{EOF} (in which the atom-field coupling is constant) with the right ones implies that the presence of intensity-dependent coupling may improve the maxima of the atom-field entanglement. Considering the effect of $f$-deformed Kerr nonlinearity, it is understood that the existence of deformed Kerr medium can enhance the amount of DEM, as the detuning parameter. While the Stark shift reduces the entanglement between the atoms and the fields.


\subsection{Atom-atom entanglement}\label{Atom-atom entanglement}
In this subsection, we study the dynamics of the DEM between the atoms (atom-atom entanglement) via definition of the EOF measure for mixed state (equation (\ref{EOF3})).
 Our  results in figure \ref{Entanglement-of-Formation} represent the time evolution of the EOF against the scaled time $\tau$ for the same parameters as in figure \ref{EOF}. The left (right) plots again correspond to $f(n)=1$ ($f(n)=\sqrt{n}$).
From figure \ref{Entanglement-of-Formation}(a) showing the resonance case, in the absence of Kerr nonlinearity and Stark shift, it is observed that the EOF has a random behavior. Comparing both left and right plots of this figure indicates that the intensity-dependent coupling can increase the maximum amount of EOF. Figure \ref{Entanglement-of-Formation}(b) which deals with effect of deformed Kerr medium, shows that this nonlinearity has no remarkable effect on the behavior of EOF, as is compared with figure \ref{Entanglement-of-Formation}(a). The effect of detuning parameter is presented in figure \ref{Entanglement-of-Formation}(c). According to this figure, it is obviously seen that, the DEM between the atoms is drastically diminished as time goes on. However, this will be faster for the case of intensity-dependent coupling. In addition, it may be noted that, in the presence of detuning parameter, the maximum value of EOF for the case $f(n) = \sqrt{n}$ is smaller than $f(n)=1$. The influence of the Stark shift in the resonance case and in the absence of the Kerr nonlinearity is depicted in figure \ref{Entanglement-of-Formation}(d). It can be seen from this figure that, in the constant coupling regime, the amount of EOF is considerably ascended especially when the time proceeds. While in the intensity-dependent coupling, the DEM between the atoms is preserved.

 \section{Summary and conclusion}\label{Summary}
Due to the fact that, the light-matter interaction in cavity QED is considered as a usual way to generate various classes of entangled states, in this paper, we have outlined a paticular nonlinear interaction between two two-level atoms and two-mode quantized radiation field in an optical cavity containing a medium with centrosymmetric Kerr nonlinearity in the presence of Stark shift, detuning parameter and intensity-dependent coupling. After suitably considering all existing interactions, the explicit form of the state vector of the entire system has been obtained. Then, we have aimed to study the dynamics of entanglement of the system consisting of the two atoms and the two-mode field. Consequently, we have planned to examine the temporal behavior of the DEM between the available subsystems, which may be observed during the interaction, through the study of the EOF measure.\\

Summing up, the main outcomes of the considered interaction model and the related numerical calculations are briefly listed as below.
\begin{itemize}

 \itemsep1em

 \item {\it Tuning the nonclassicality indicators:}  It is illustrated that the amount of the EOF measure can be tuned by choosing the nonlinear parameters related to the atom-field system, suitably.
   \item {\it Intensity-dependent coupling:} Presented results show that intensity-dependent coupling (which is considered by the function $f(n) = \sqrt{n}$) may in general enhance the EOF.
  \item {\it Deformed Kerr medium:} Paying attention to the related results implies that the deformed Kerr medium (which is distinguishable from Kerr medium through the nonlinearity function $g(n) = 1/\sqrt{n}$) has an obvious role and noteworthy effect in improving the entanglement.
   \item {\it Detuning parameter:} Looking deeply at the obtained results shows that the detuning parameter can ameliorate the atom-field entanglement. While this parameter reduces the DEM between the atoms drastically.
   \item {\it Stark shift:} From the numerical results, it is revealed that, in the constant coupling regime, the amount of the DEM is apparently decreased in the presence of Stark shift. Although, in the case of intensity-dependent coupling, the Stark shift can increase the DEM when it is compared with the constant coupling regime.
 \end{itemize}
%


\section*{References}


\providecommand{\newblock}{}

 \end{document}